\documentstyle[epsfig]{elsart}
\begin{document}
\begin{frontmatter}
\title{Limits on the low-energy antinucleon annihilations from the 
Heisenberg principle} 
\author[Brescia]{A.~Bianconi.} 
\address[Brescia]{Dip. di Chimica e Fisica per 
l'Ingegneria e per i Materiali, 
Universit\`{a} di Brescia and
INFN, Sez. di Pavia, Italy} 

\begin{abstract}
Here a short synthesis is presented of the work, developed 
in the last two years 
by the Brescia Collaboration, on the phenomenology  
of antinucleon-nucleon and antinucleon-nucleus 
annihilation at small momenta (below 300 MeV/c in 
the laboratory), with special stress on the role of general 
principles. 
\end{abstract} 
\end{frontmatter} 

Our work\cite{n1} in the last two years was mainly 
devoted to the study of the strong shadowing characterizing 
$\bar{p}$ annihilations onlight nuclei. 
The nuclear shadowing is evident in the fact that the 
$\bar{p}p$ total annihilation cross sections are larger 
than the corresponding $\bar{p}$D and $\bar{p}^4$He ones 
for $\bar{p}$ momenta in the laboratory 
$k$ $<$ 70 MeV/c\cite{obe}. 
In previous works we used the word ``inversion''. 
A related phenomenon was studied in antiprotonic 
atoms\cite{wid,bat90,wyc,cp1} and also pionic atoms\cite{fs}. 

Since ``shadowing'' can be defined as 
the amount of departure from the Impulse Approximation, 
to quantify its real presence we need first to produce  
an impulse approximation estimation of the measured nuclear data. 
Not to overrate the shadowing effects, 
two more points must be taken into account: (1) a 
correct estimation of Coulomb enhancement effects, and 
(2) center of mass effects. When data are represented 
with respect to the center of mass momentum\cite{n1}, the shadowing 
effect is smaller, although apparently all the 
available nuclear annihilation cross sections become quite 
similar at $k_{cm}$ $<<$ 100 MeV/c. To fully understand 
the relevance of this similarity, we remind that at low energies 
Coulomb effects are expected to be quite strong and enhance 
$\bar{p}^4$He annihilation rates 
about twice with respect to $\bar{p}p$ ones. 
This traditional estimate of Coulomb effects\cite{ll1} is 
based on the approximation of pointlike 
particles. Within an optical potential framework we 
re-calculated these 
Coulomb corrections, taking into account the finite size of the 
$\bar{p}$ and nuclear charge distributions. 
One of the results\cite{n1} was that the advantage of 
the Helium charge with respect to the Hydrogen case 
was not that large, in the laboratory frame. 
This advantage is however stronger with respect to center of 
mass momenta. The role of the reference frame can be understood 
taking into account that 
low energy exoenergetic reaction cross sections are supposed to 
be roughly $\sim$ $1/k_{cm}$ for neutral projectiles, and 
$\sim$ $1/k_{cm}^2$ for charged projectiles. E.g.,  
$k_{lab}$ $=$ 100 MeV/c means $k_{cm}$ $\approx$ 100 MeV/c 
for $\bar{p}^4$He, and $k_{cm}$ $=$ 50 MeV/c for $\bar{p}p$. 


\begin{figure}[htp]
\begin{center}
\mbox{
\epsfig{file=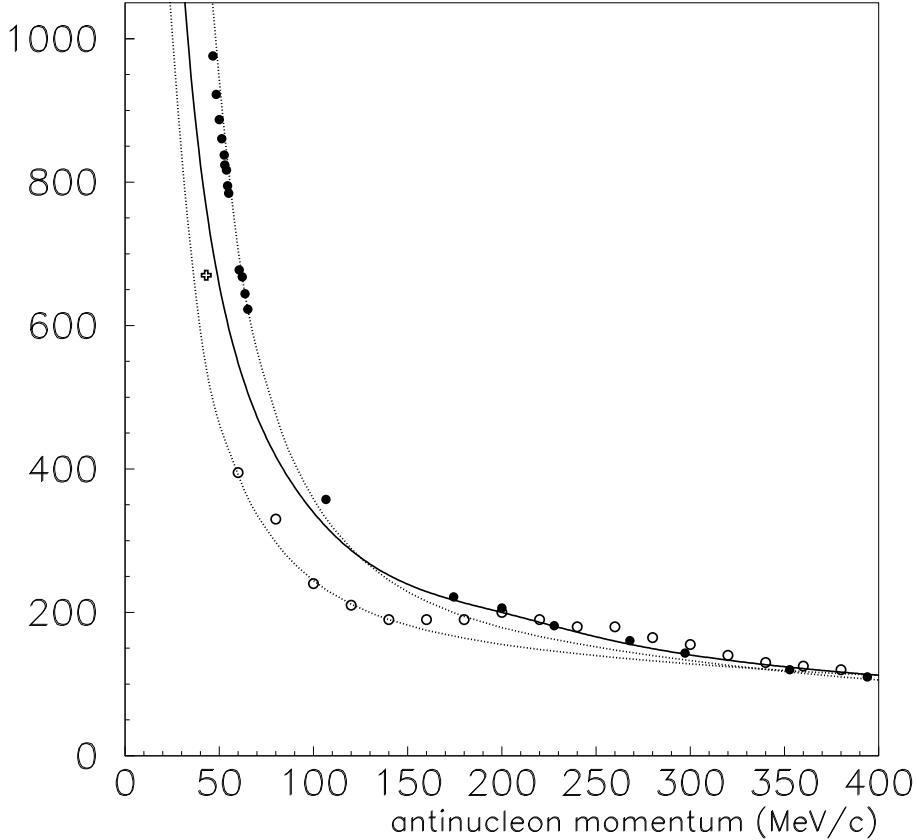,width=0.9\linewidth}}
\end{center}
\caption[]
{\small\it \label{fig1}
Antineutron (empty circles) and antiproton (full circles)
total annihilation cross sections (mb)
measured by the Obelix experiment\cite{feli,obe}. 
The empty crosses reproduce the two low-energy
$\bar{n}p$ total annihilation points measured in \cite{mutch}.
Error bars are not reported here.
The two dotted lines 
correspond to optical potential fits (see text for details).
The solid line represents the $\bar{p}p$ annihilation
cross section after Coulomb effects have been subtracted. 
The lower energy part of this curve has been calculated
by extrapolating the optical potential fit of the $\bar{p}p$
data and by removing the electrostatic part
of the potential. For $k$ $>$ 30 MeV/c the Coulomb effects have
been subtracted from the actual $\bar{p}p$ points, not from
the potential fit (for this
reason the solid curve is larger than the dotted curve
for $k$ $>$ 130 MeV/c). }
\end{figure}

For calculating a PWIA $\bar{p}$-nucleus annihilation rate, 
we need $\bar{p}p$ and $\bar{p}n$ annihilation rates in a suitable 
momentum range. The latter have been supposed to be equal to the 
$\bar{n}p$ measured annihilation rates\cite{feli} (see fig.3).  
We have subtracted Coulomb effects 
from the measured $\bar{p}p$ annihilation 
rates, leading to the ``uncharged $\bar{p}p$ annihilation 
cross section'' shown by the solid line in fig.1. After calculating 
the total nuclear annihilation cross section, the result has been 
rescaled by the $nuclear$ enhancement Coulomb factor. 
This procedure is 
needed for the following reasons. When a $\bar{p}$ 
annihilates on a nucleus, the Coulomb forces have two effects: 
(1) a focusing of the $\bar{p}$ wave in the reaction region, and 
(2) an increase of the $\bar{p}$ kinetic energy. Both effects 
take place on an atomic scale $\sim$ $r_B$ $>>$ $R_{nucleus}$. 
So $\bar{p}p$ and $\bar{p}n$ processes are equally ``Coulomb 
affected'' when the proton and neutron are bound to the same 
nucleus. 

Last, the PWIA calculated annihilation cross section 
has been renormalized to the measured value at $k$ $\approx$ 
350 MeV/c. This permits to remove the eclipse effect from 
the calculations. This effect is well known, and reduces 
the annihilation rates by a slowly energy dependent factor at 
all energies, leading to a reaction rate proportional to 
$A^{2/3}$. This effect can be considered a component of the 
shadowing, but clearly it is not what we are interested in. 
In fig.2 we show the results of the IA fit on deuteron. We 
produce three curves. One takes into account the real deuteron 
composition, the other two assume a deuteron composed by two 
neutrons or two protons (in both cases with total deuteron electric 
charge 1). The comparison between the three curves, and between 
the solid one and the data suggests that (1) not to take into 
account the actual proton/neutron composition of the nucleus 
introduces large errors for $k$ $<$ 200 MeV/c, and (2) remarkable 
shadowing is anyway present for $k$ $<$ 100 MeV/c. 

\begin{figure}[htp]
\begin{center}
\mbox{
\epsfig{file=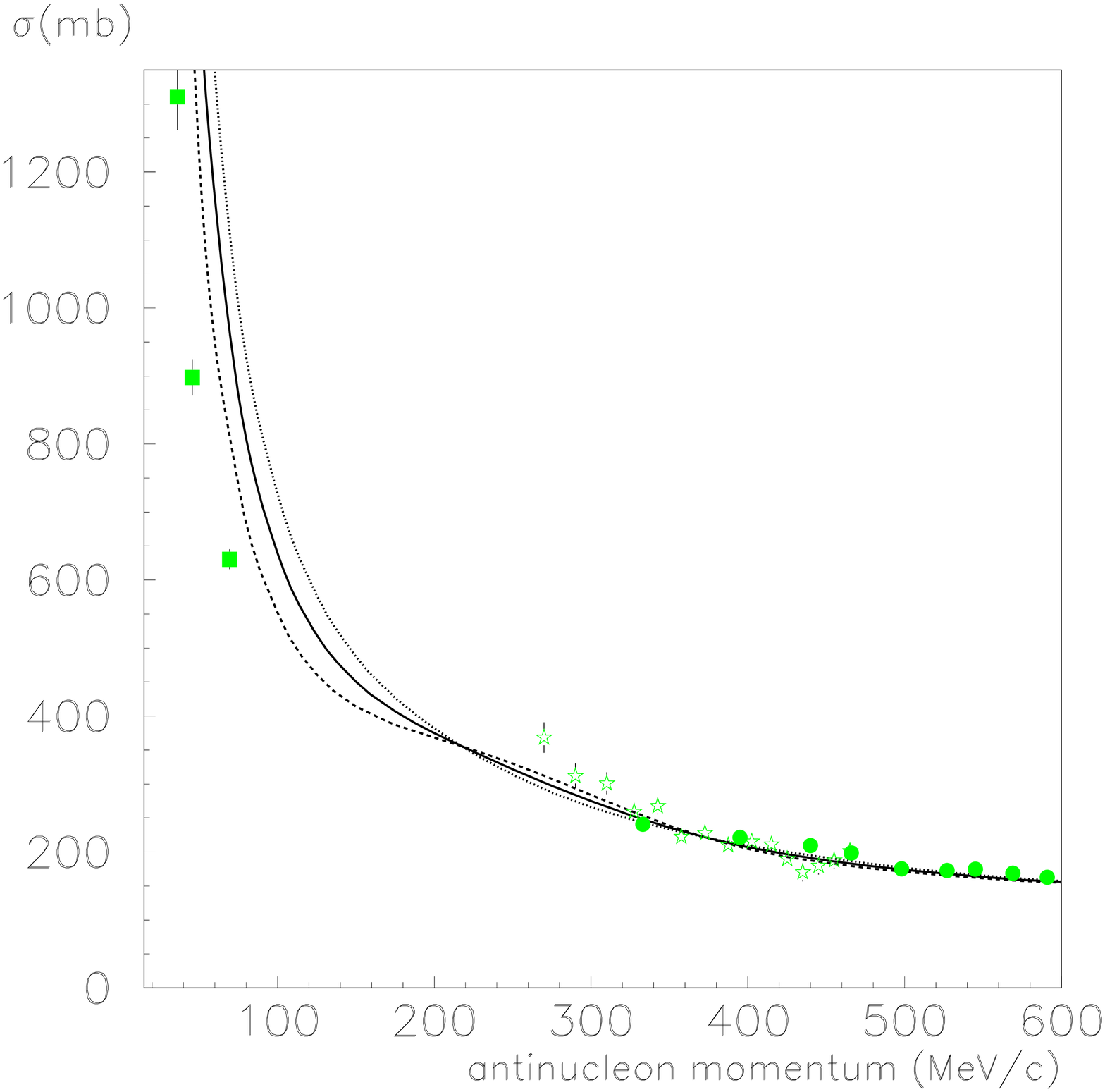,width=0.9\linewidth}}
\end{center}
\caption[]
{\small\it \label{fig2}
PWIA calculation of the $\bar{p}$-Deuteron total
annihilation cross sections, together with data points
taken from references \cite{obe} (full squares),
\cite{kalo80} (empty stars), \cite{bizz80} (full circles).
Continuous
line: full PWIA calculation with Coulomb correction and
renormalization of the curve to the point at 340 MeV/c.
Dashed line: the Deuteron is supposed to be composed
by two neutrons (with overall nuclear charge Z=1).
Dotted line: the Deuteron is supposed to be composed
by two protons (with overall nuclear charge Z=1).
}
\end{figure} 

Rather independently of the mechanism underlying the 
annihilation process, it had been 
previously demonstrated that in the framework of the 
multiple scattering theory\cite{wyc}, of variational 
methods\cite{wyc} and of optical potential 
treatments\cite{n1,pro,kg96,bat90} one can predict such shadowing 
effects. The effect seems to be present also in pionic 
atoms\cite{fs}. 

The fact that different methods lead to similar results
suggested us to investigate the problem from a more 
general and qualitative, although less precise, point 
of view. 
In our work we have shown that 
due to the quantum uncertainty principle the 
$\bar{n}$-nucleus cross sections 
should be almost $A$-independent, apart for fluctuations due 
to nuclear surface effects. Consequently the $\bar{p}$-nucleus 
cross sections should depend on the target because of its 
electric charge only. The underlying argument is that 
most of the existing models\cite{dov1} 
or phenomenological analyses\cite{bonn1,brue91,aar} 
establish that the annihilation 
process takes place when the centers of mass of the 
antinucleon and of the target nucleus 
are at a relative distance $d$ such that 
$R_{nucleus}$ $<$ $d$ $<$ $R_{nucleus}+\Delta$, where 
$\Delta$ $\sim$ 1 fm (or smaller, depending on 
the model) and $\Delta$ does not depend too much on the target. 
So the annihilation is equivalent to 
a measurement of the projectile-target relative distance 
with uncertainty $\Delta$ $<$ 1 fm, and this measurement is 
incompatible with a relative momentum $<<$ 200 MeV/c. 

We distinguish between two classes 
of nuclear reactions. On one side, inelastic reactions 
where the entire nucleus is involved, as in compound nucleus 
reactions, but the underlying projectile-nucleon 
processes are elastic (e.g. neutron induced nuclear 
reactions). In this case the characteristic reaction 
region coincides approximately with the target nucleus. 
Then the uncertainty $\Delta$ coincides approximately 
with the nuclear radius. On the other side, we find 
reactions where a strong inelasticity is present at the 
projectile-nucleon level. In this case reactions deep  
inside the nuclear volume are rare, the reaction region 
is a shell at the surface of the target nucleus, with 
thickness $\Delta$, and $\Delta$ is approximately the 
same for all the possible targets. 
\begin{figure}[htp]
\begin{center}
\mbox{
\epsfig{file=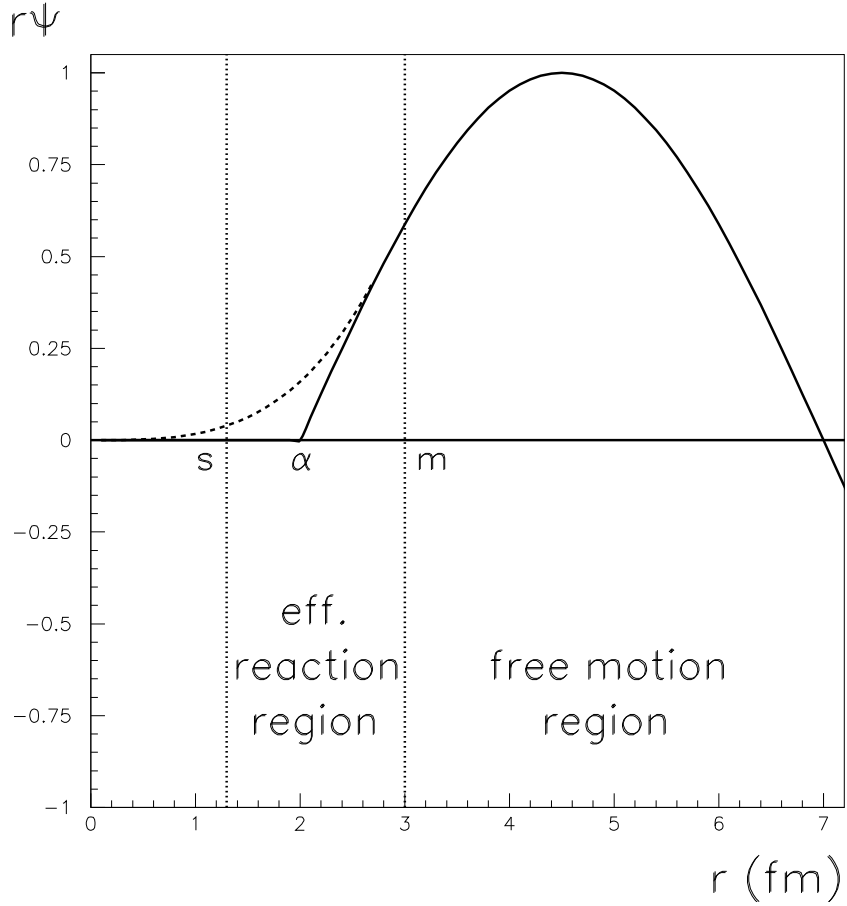,width=0.9\linewidth}}
\end{center}
\caption[]
{\small\it \label{fig5}
S-wave projectile wavefunction $\chi(r)$ $=$
$r\cdot\Psi(r)$ in the reaction region and around it. In
the free motion region $\chi$ (continuous line) has the form
$sin[k(r-\alpha)]$. Here $\alpha$ is assumed real
for graphical reasons. Reactions are possible for
$r$ $<$ $m$, a matching radius ($R_{nucleus}+\Delta$ in the 
text), however cospicuous
damping of the projectile wavefunction is reached in the
``effective reaction region''. The (approximate) lower
limit $r$ $=$ $s$
of this region is defined by the condition
$\chi(s)$ $<<$ $\chi(m)$. What happens for
$r$ $<$ $s$ is not relevant. The parameter $\Delta$ used
in the text coincides with $\vert s-m \vert$ in this
example. In the matching point $r$ $=$ $m$
the free motion wave
must join regularly with the internal wave (dashed line).
The scattering length $\alpha$ represents the ``virtual source''
of the free motion wave, i.e. the position where the extrapolation
of the free motion wave reaches the $r$-axis.
The logarithmic derivative $\chi'/\chi$
in the matching point is clearly inversely proportional to the
size $\Delta$ of the effective
reaction region. For a fixed value of $\chi$
at the matching point $m$, the larger the derivative
$\chi'$ in $m$ the larger the peak value of the wavefunction in
the free motion region. In
this example the wavelength is 10 fm, and is much larger
than $\Delta$, with the consequence that the peak value of
the free motion function is much larger than its value at the
matching point. With a wavelength of e.g. 0.5 fm
(i.e. $<<$ $\Delta$) this would not be true.
}
\end{figure}

The consequence of the limitations imposed by the 
uncertainty principle is that for antinucleon 
momenta $k$ $<<$ $1/\Delta$ the total reaction cross section 
becomes much smaller than its possible unitarity limit. 
In the limit of very small momenta 
this is also established by the well known\cite{ll1} 
low energy limit for the phase shifts: 
$\delta_l$ $\propto$ $k^{2l+1}$ for $k$ $\rightarrow$ 0. 
The unitarity limit is reached when a 
partial wave is completely absorbed in the reaction 
process, which means $exp(i\delta_l)$ $=$ 0, i.e. 
$Im(\delta_l)$ $=$ $\infty$, so the unitarity limit 
cannot be attained at small enough $k$. Uncertainty 
considerations suggest that for $k$ $>>$ $1/\Delta$ 
it is possible, for strong enough reactions, to 
saturate the unitarity limit, while for $k$ $<<$ 
$1/\Delta$ we are in the situation where $\delta_l$ 
$=$ $O(k^{2l+1})$, whatever the strength of the reaction. 

On the ground that the projectile wavefunction $\chi$ 
$\equiv$ $r\Psi(r)$ 
is completely damped within a range $\Delta$ (i.e 
$\vert \chi\vert$  
is large for $r$ $>$ $R_{nucleus}+\Delta$ and very small 
for $r$ $<$ $R_{nucleus}$) 
it is straightforward to demonstrate that for the 
scattering length $\alpha$ we have (approximately):

$Im(\alpha)$ $\approx$ $-\Delta$, 

$Re(\alpha)$ $\approx$ $+R_{nucleus}$. 

\noindent
These results derive from the observation that 
the $\chi$ damping requirement implies, 
for the logarithmic derivative of the projectile wavefunction, 
$\vert \chi'/\chi\vert$ $\approx$ 
$1/\Delta$. This is an obvious geometrical consequence 
of the damping requirement $\chi(R-\Delta)$ $<<$ $\chi(R)$, 
but in more physical terms it is a consequence of the 
uncertainty principle. 
When the absolute value of the logarithmic 
derivatives of the free motion wavefunction  
$\vert \chi'/\chi\vert_{r=R_{nucleus}+\Delta}$ $=$ 
$\vert k\cdot cotg\{k(R_{nucleus}+\Delta-\alpha)\}\vert$ 
is matched with the corresponding quantity for 
the wavefunction in the annihilation region 
$\vert \chi'/\chi\vert_{r=R_{nucleus}+\Delta}$ $=$ 
$1/\Delta$, the previous values for the scattering length 
are obtained in the limit $k$ $<<$ $1/\Delta$. 
An illustration of this is given in fig.3. 

A paradoxical consequence is that 
a smaller $\Delta$ corresponds to what would be a 
stronger reaction at large energies, so that at low 
energies ``stronger'' interactions lead to a smaller 
reaction rate. 

The above values of course are deduced from approximate 
equations, so they represent just estimates, however 
they suggest that the antineutron 
annihilation cross sections should  
not show a $systematic$ increase with the target mass 
number $A$. Such an increase could be present for 
antiproton annihilations, but because of Coulomb effects 
only. When going to any specific target nucleus, 
non-systematic effects will be present, 
related with the structure of the nuclear surface. 

The exposed mechanism has an interesting consequence in 
the case of optical potential analyses: an increase of the 
strength of the imaginary part of the optical potential can 
lead to a decrease of the consequent reaction rate at 
small momenta. In the above 
language, an increase 
in the potential strength leads to a decrease in the 
size parameter $\Delta$, since the absorption of the 
projectile wavefunction takes place in a shorter range. 
Also modification of other parameters 
(radius, diffuseness, etc) leads to consequences that 
are not necessarily the most obvious ones. A relevant example is 
given next. 

In fig.1 we have shown, together with data, two optical 
potential fits. The $\bar{p}p$ fit uses a slight modification 
of an optical potential used by another group\cite{brue91} 
to fit elastic data 
at 200-600 MeV/c. It is a Woods-Saxon shape  
with $V_I$ $=$ $-$8000 MeV, $V_R$ $=$ $-$46 MeV, $R_I$ $=$ 
0.52 fm, $R_R$ $=$ 1.89 fm, $a_I$ $=$ $a_R$ $=$ 0.15 fm. 
The Coulomb potential is the potential of a spherical 
charge distribution with radius $R_c$ $=$ 
$\sqrt{R_p^2+R_{\bar{p}}^2}$ $=$ 1.25 fm. 
The interesting point is that the $\bar{n}p$ fit is obtained 
either by increasing $V_I$ to 14000 MeV, or by increasing 
$R_I$ to 0.75 fm (in addition to removing Coulomb effects). 
In both cases a ``more effective'' 
annihilating potential leads to a smaller cross section. 

It is interesting to observe that the fit by an $energy$  
$independent$ optical potential can reproduce very 
well the $\bar{p}p$ annihilation cross section, and follows 
the ``average'' trend of the $\bar{n}p$ one, but is unable 
to reproduce the broad peak that is present at 150-350 MeV/c 
over this ``background''. At very low momenta the $\bar{n}p$ 
cross section rises again, reaching the ``uncharged $\bar{p}p$'' 
one, represented by the solid line in fig.1. So we speak of 
a ``regular background'', dominant in the $\bar{p}p$ 
case, also relevant in the $\bar{n}p$ case, that can be 
reproduced by an energy independent optical potential, 
and of a ``gap/peak'' structure that corresponds to a 
more complicate physics than pure absorption. This structure 
is evidently characteristic of the isospin-1 channel. 

To conclude, we may say that the nuclear annihilation rates at 
low energies can be partly explained by the difference between 
$\bar{p}p$ and $\bar{p}n$ interactions. On the other side, a 
strong shadowing is present. We think this shadowing to be 
due to the ``inversion'' behavior of the strongly absorbing 
processes at low energies, inversion behavior that can be 
justified as a manifestation of the Heisenberg uncertainty 
principle. This also has the consequence that probably, despite 
appearances, $\bar{n}p$ interactions are $stronger$ than 
$\bar{p}p$ ones at low energies. We also stress that the ratio 
between $\bar{n}p$ and $\bar{p}p$ annihilation rates 
presents an oscillating behavior that cannot 
be justified by regular absorption only.

\end{document}